\documentclass[twocolumn,aps,pra,amsmath,amssymb,showpacs]{revtex4-1}
\usepackage[latin9]{inputenc}
\usepackage{color}
\usepackage{amsmath}
\usepackage{amssymb}
\usepackage{graphicx}
\usepackage{esint}
\usepackage[unicode=true,pdfusetitle,
 bookmarks=true,bookmarksnumbered=false,bookmarksopen=false,
 breaklinks=false,pdfborder={0 0 0},pdfborderstyle={},backref=false,colorlinks=true]
 {hyperref}
\hypersetup{
 colorlinks,linkcolor=red,citecolor=blue}
\usepackage{breakurl}

\makeatletter
\usepackage{epsfig}\usepackage{amsfonts}

\makeatother

\makeatother

\begin{document}

\title{``M\"obius'' Microring Resonator}

\author{Xin-Biao Xu}

\address{Key Laboratory of Quantum Information, University of Science and
Technology of China, CAS, Hefei, Anhui 230026, China. }

\address{Wuhan National Laboratory for Optoelectronics, Huazhong University
of Science and Technology, Wuhan 430074, People Republic of China}

\author{Lei Shi }
\email{lshi@hust.edu.cn}

\address{Wuhan National Laboratory for Optoelectronics, Huazhong University
of Science and Technology, Wuhan 430074, People Republic of China}

\author{Guang-Can Guo}

\address{Key Laboratory of Quantum Information, University of Science and
Technology of China, CAS, Hefei, Anhui 230026, China. }

\author{Chun-Hua Dong}

\address{Key Laboratory of Quantum Information, University of Science and
Technology of China, CAS, Hefei, Anhui 230026, China. }

\author{Chang-Ling Zou}
\email{clzou321@ustc.edu.cn}

\address{Key Laboratory of Quantum Information, University of Science and
Technology of China, CAS, Hefei, Anhui 230026, China. }

\address{National Laboratory of Solid State Microstructures, Nanjing University,
Nanjing 210093, China.}
\begin{abstract}
A new category of optical microring resonator, which is analogous
to the M\"obius strip, is proposed. The \textquotedblleft M\"obius\textquotedblright{}
microring resonator allows the conversion between modes with different
polarizations in the ring and light must circulate two cycles to be
converted back to the original polarization status, which is similar
to a M\"obius strip. This topology structure of polarization leads to
the free spectral range be half of that corresponds to the cavity
round trip. The eigenmodes of this microring are hybridizations of
different polarizations and the breaking of the rotation invariance
of the ring makes the transmission be related with the polarization
of input light. Our work opens the door towards the photonic devices
with nontrivial mode topology and provides a novel way to engineering
photonic structures for fundamental studies. 
\end{abstract}
\maketitle
Over the past decades, photonic integrated circuits (PICs) have been
extensively studied \cite{Soref2006IJSTQE,Politi2008S,Politi2009IJSTQE}.
The high refractive index contrast between the waveguides and their
cladding allows for the realization of compact optical devices. Besides,
because of the CMOS-compatible materials, the photonic circuits can
be fabricated by the well-developed processes for integrated electronic
circuits. All of these advantages make PICs be of great potential
for scalable classical and quantum information processing \cite{Zhang2016O,Sun2015N,Matthews2009NP,Khasminskaya2016NP}.
Among the various components of PICs, integrated optical microcavities
are crucial \cite{Vahala2003N} and different kinds of microcavities
have been demonstrated over the past years \cite{Bogaerts2012LPR,Jiang2017S,Poon2018NP}.
Owing to the advantages of high quality factor, small mode volume
and reconfigurability, microring resonators made by a closed uniform
waveguides are used in many realms, including filter \cite{Little1997JLT},
ultrasensitive sensor \cite{Foreman2015AOP,Wan2018PR}, nonlinear
optics \cite{Xiong2012NJP,Heylman2016NP,Guo2018PRL}, and optomechanics
\cite{Bagheri2011NN,Bagheri2013PRL}.

\begin{figure}
\includegraphics[width=1\columnwidth,height=0.25\columnwidth]{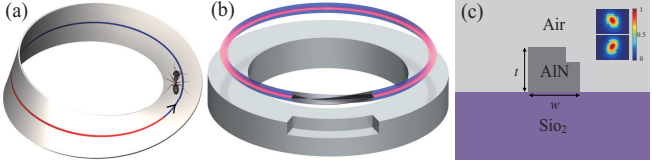}\caption{\label{fig1}(a) The M\"obius strip. The red and the blue lines mark
two faces of the original strip (b) The microring resonator, which
is analogous to the M\"obius strip. There is a region with a slot at
the corner, which induces the coupling between two polarization modes.
The red and the blue circles upon the microring represent two polarization
modes in the waveguide. (c) The cross section and mode fields of two
eigenmodes of the polarization rotator in the MMR.}
\end{figure}

By twisting one end of an ordinary waveguide through $180^{\circ}$
and making the ends join together, a new resonator similar to the
M\"obius strip will be created \cite{Starostin2007Nm}. The M\"obius strip
is famous for its nontrivial topological structure. As is shown in
Figure$\,$\ref{fig1}(a), a M\"obius strip has only one surface, an
ant traveling along the strip can come back to the same point after
two cycles, and thus the length of roundtrip path is double of that
of the original strip. In recent years, a series of studies about
Mobius strips emerge in optics \cite{Freund2010OL,Bauer2016PRL,Garcia-Etxarri2017AP},
but it\textquoteright s very challenging to fabricate a photonic structure
with such a complex geometric topology \cite{Kreismann2018,Li-Ting2017JO,Yin2017LPR}. 

In this paper, we propose a new category of microring resonator with
non-trivial mode polarization properties. By properly engineering
the microring with a polarization rotator (PR) \cite{Watts2005OL,Fukuda2008OE,Wang2008JOSAB,Chen2011OL,Xiong2013OE},
light must circulate two cycles to be converted back to the original
status. Therefore, the free spectral range (FSR) of the resonator
is approximately $1/2$ of that for a traditional microring without
PR, which makes the resonator be analogous to a M\"obius strip \cite{Bauer2015S}.
Furthermore, the effective optical path of the resonator can be $N$
times of the optical circumference for a generalized $N$-fold \textquotedblleft M\"obius\textquotedblright{}
microring resonator (MMR). Such a new kind of resonator promises the
applications for sensing and polarization analysis.

The microring resonator proposed to realize the ``M\"obius'' topology
in the virtual dimension of modes instead of real geometry space is
shown in Fig.$\,$\ref{fig1}(b), where the two polarization states
of the light in the ring are analogous to the two surfaces of the
original strip. As we know, due to the structural characteristics
of the integrated waveguide, rectangular waveguides exhibit strong
polarization-maintaining properties with well-defined TE (H-polarized)
and TM (V-polarized) polarized modes supported. When etched corner
is introduced at a section of nonuniform waveguide, the H- and V-polarization
modes in rectangular waveguide are not the eigenmodes of the etched
waveguide anymore, which are shown in the inset of Fig.$\,$\ref{fig1}(c),
and the H- and V-polarization modes can convert to each other in the
etched waveguide, so it serves as a PR. When the conversion rate between
the two polarization modes approaches unit, the H-polarization mode
traveling in the microring will become V-polarization mode after one
cycle of propagation and the V-polarization mode will be converted
back to the H-polarization mode after another cycle of propagation.
So the resonator is similar to a M\"obius strip and the two polarization
modes corresponding to the two surfaces of the original strip. In
this letter, we choose the Aluminum Nitride (AlN) microring for analyses,
since AlN possesses excellent linear and nonlinear optical properties
\cite{Xiong2012NJP,Jung2013OL,Guo2016PRL}. The thickness and width
of microring are $t=650\,\mathrm{nm}$ and $w$, respectively. The
working wavelength is around $\text{\ensuremath{\lambda}=}1550\,\mathrm{nm}$
with the refractive index about $2.1$. Instead of directly solving
the modes in the MMR, we start with the analyses of the propagating
light in it.

\begin{figure}[h]
\includegraphics[width=1\columnwidth,height=0.71\columnwidth]{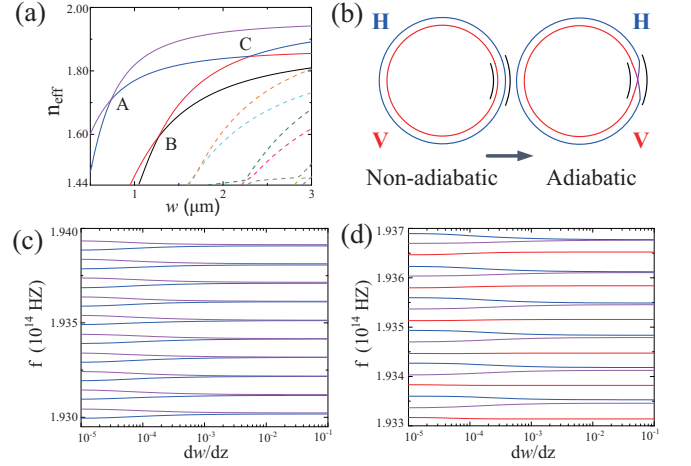}\caption{\label{fig2}(a) The effective refractive index of different modes
versus waveguide width of the PR. (b) The mode topology of the MMR
of nonadiabatic and adiabatic mode conversion. (c) The spectrum of
MMR for different waveguide width varying rate $dw/dz$ when there
is only one PR in the ring with the waveguide width around the avoid
crossing point $A$ shown in Fig. \ref{fig2}(a). (d) The spectrum
of MMR for different waveguide width varying rate $dw/dz$ when there
are two PRs in the ring with the waveguide width around the avoid
crossing points $B$ and $C$, respectively.}
\end{figure}

For the light propagating along the PR, the corresponding mode effective
refractive index $n_{\mathrm{eff}}$ for different waveguide width
of the PR is shown in Fig.$\,$\ref{fig2}(a). Due to the etched corner
induced mode coupling, there are several avoid crossing regions, i.e.
$A$, $B$, and $C$. In these regions, the coupled-mode equations
for H- and V-polarized modes read
\begin{equation}
\frac{d}{dz}\left[\begin{array}{c}
E_{H}\\
E_{V}
\end{array}\right]=\left[\begin{array}{cc}
-ik_{0}n_{H} & -ig\\
-ig & -ik_{0}n_{V}
\end{array}\right]\left[\begin{array}{c}
E_{H}\\
E_{V}
\end{array}\right],\label{eq:1}
\end{equation}
where light propagates along the $z$-direction, $k_{0}=2\pi/\lambda$,
$n_{H(V)}$ is the $n_{\mathrm{eff}}$ of H- (V-) polarized mode,
$g$ is the coupling strength. The coupling leads to the minimum propagation
constant difference between normal modes is $2g$. To realize a broadband
and robust PR, the waveguide width $w$ along the $z$-direction of
the PR is varying slowly. According to the Landau-Zener tunneling
theory \cite{Zener1932PRSLA,Khomeriki2005PRL}, the maximum conversion
rate can be estimated by
\begin{equation}
\eta\approx1-e^{-2\pi g^{2}/\left|\kappa\frac{dw}{dz}\right|},\label{eq:Landau-Zener}
\end{equation}
where $dw/dz$ is the varying rate of the waveguide width and $\kappa=k_{0}d(n_{H}-n_{V})/dw$.
Therefore, the evolution of the H- and V-polarized light after one
round-trip can be described by 
\begin{equation}
\left[\begin{array}{c}
E_{H}^{'}\\
E_{V}^{'}
\end{array}\right]=T\left[\begin{array}{c}
E_{H}\\
E_{V}
\end{array}\right],\label{eq:3}
\end{equation}
with the transfer matrix
\begin{equation}
T=\left[\begin{array}{cc}
e^{-i\varphi_{H}} & 0\\
0 & e^{-i\varphi_{V}}
\end{array}\right]\left[\begin{array}{cc}
\sqrt{1-\eta} & \sqrt{\eta}\\
-\sqrt{\eta} & \sqrt{1-\eta}
\end{array}\right].\label{eq:4}
\end{equation}
We can further solve the eigenmodes by the $T$, and obtain the spectrum
of an MMR with different $dw/dz$. In Fig. \ref{fig2}(c), we focus
on the coupling of fundamental H- and V-polarized modes with $w$
around 740 $nm$ (avoid-crossing region $A$) and the effective radius
of the ring is 200 $\mu m$. The slot size is fixed to $50nm\times50nm$,
which induces the coupling strength $g=10^{-3}k_{0}$ and $\kappa=0.31k_{0}$
$\mathrm{\mu m^{-1}}$. In the following analyses, the PR is treated
as a point and we neglect the change of phase in the coupling region
when calculating the roundtrip phase for H- or V-polarized light.

When $dw/dz$ is large, $\eta$ is small as indicated by Eq.$\,$(\ref{eq:Landau-Zener}).
In Fig.$\,$\ref{fig2}(c), for $dw/dz\rightarrow0.1$, the conversion
is non-adiabatic, the propagation of H- and V-polarized modes are
independent and thus there are two independent families of spectra
for two polarization modes, just corresponds to that in traditional
microrings. However, by reducing the geometry parameter $dw/dz\rightarrow10^{-5}$,
the conversion will be adiabatic and $\eta$ approaches $1$. In this
case, the two separated optical paths of two polarization modes combine
to one path and therefore the effective length of the path is double.
As a result, the eigenmodes of MMR are hybridizations of H- and V-polarized
light and the FSR becomes half of that when the conversion is non-adiabatic.
The mode topology when conversion changes from non-adiabatic to adiabatic
is shown in Fig.$\,$\ref{fig2}(b). 

In Fig.$\,$\ref{fig2}(d), we generalize the MMR to three mode families
by including two PRs based on the avoid-crossing regions $B$ and
$C$, respectively. Similarly, we found that when the conversion in
PRs are non-adiabatic, the propagation of three modes are independent
and there are three sets of independent spectra. When the conversion
is adiabatic, the eigenmodes of the ring will be a hybridization of
these three modes and the length of light path approximatively becomes
$3$ times of that when the conversion is non-adiabatic. The concept
of \textquotedblleft M\"obius\textquotedblright{} mode topology can
be further developed by introducing higher order modes in the ring.
When there are $N$ adiabatic PRs in the ring with $N$ cascaded avoid
crossings as shown in Fig. \ref{fig2}(a), the light path will become
$N$ times of the circumference for the $N$-fold MMR.

\begin{figure}[h]
\includegraphics[width=1\columnwidth,height=1.05\columnwidth]{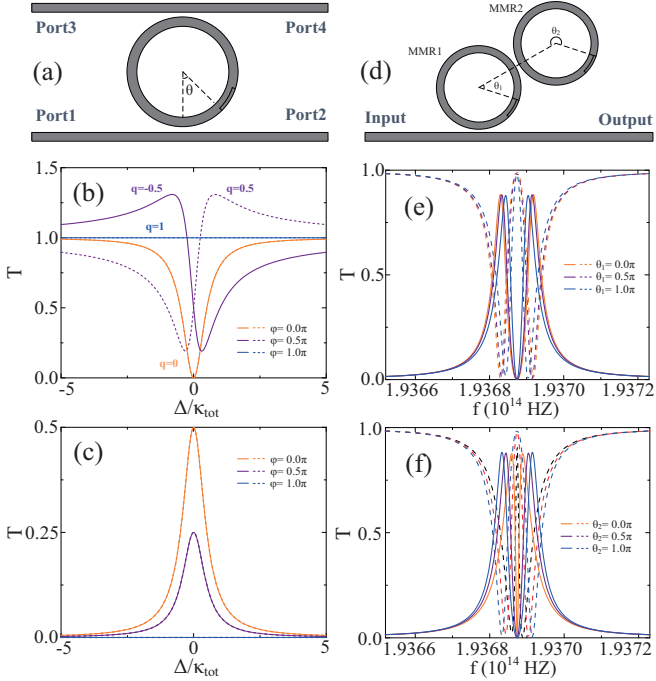}\caption{\label{fig:3}(a) The schematic of add-drop MMR. The light is input
at port 1. (b) and (c) are the transmission spectra of two polarization
modes monitored at port 2 and 3 respectively with different phase
$\varphi$ when the amplitude ratio of the $E_{V}$/$E_{H}$ is fixed
at 1. (d) The schematic of coupled MMRs. (e) and (f) are the transmission
spectra of two polarization light with different $\theta_{1}$ ($\theta_{2}=\pi$)
and $\theta_{2}$ ($\theta_{1}=0.5\pi$) respectively and the amplitude
ratio of the $E_{V}$/$E_{H}$ is 0. The dash lines are the results
of H-polarized light and solid lines are the results of V-polarized
light.}
\end{figure}

In practical applications, we use adjacent waveguides to couple with
the cavity evanescently and the waveguide width is around the avoid-crossing
region $A$, as depicted in Fig.$\,$\ref{fig:3}(a). Because the
eigenmodes of the MMR are hybridizations of two polarization modes,
the cavity field will couple with the H- and V-polarized modes in
the waveguide simultaneously. The Hamiltonian of the system can be
written as \cite{Scully1997}
\begin{eqnarray}
H & = & -\Delta a^{\dagger}a+\sum_{j=1,2}\left(\sqrt{\kappa_{H,j}}E_{H,j}a^{\dagger}\right.\nonumber \\
 &  & \left.+\sqrt{\kappa_{V,j}}E_{V,j}e^{i\varphi_{j}}a^{\dagger}+h.c.\right)
\end{eqnarray}
where $a$ is the bosonic operator of the MMR cavity mode, $\Delta$
is the external probe laser frequency detuning, $\kappa_{H(V),j}$
are external coupling strength between the cavity mode and $j$-th
input waveguide of H(V)-polarized mode, $E_{H(V),j}$ are the corresponding
input fields, and $\varphi_{j}$ is the phase difference between two
polarization modes. Due to the PR in the ring breaks the rotation
invariance of the ring, the phase $\varphi$ also depends on the relative
position of the PR. If we rotate the ring by an angle $\theta$, because
of the birefringence effect, we will have $\varphi^{'}=\varphi+k_{0}R\theta(n_{V}-n_{H})$,
where $R$ is the radius of MMR. At steady state, the cavity field
is
\begin{equation}
a=\sum_{j=1,2}\frac{\sqrt{\kappa_{H,j}}E_{H,j}+e^{i\varphi_{j}}\sqrt{\kappa_{V,j}}E_{V,j}}{\Delta+j\kappa_{tot}/2},\label{eq:6}
\end{equation}
where $\kappa_{tot}=\underset{j=1,2}{\sum}\left(\kappa_{H,j}+\kappa_{V,j}\right)+\kappa_{i}$
is the total loss of the cavity and $\kappa_{i}$ is the intrinsic
loss. When light is incident only from port 1, the transmission at
port 2 is $E_{out2}=(E_{H\mathit{1}}-j\sqrt{\kappa_{H\mathit{1}}}a)\overrightarrow{e}_{H}+(E_{V\mathit{1}}e^{i\varphi_{1}}-j\sqrt{\kappa_{V\mathit{1}}}a)\overrightarrow{e}_{V}$
and the transmission at port 3 is $E_{out3}=-j\sqrt{\kappa_{H\mathit{2}}}a\overrightarrow{e}_{H}-j\sqrt{\kappa_{V\mathit{2}}}a\overrightarrow{e}_{V}$.
These equations indicate that the transmission not only depends on
the detuning but also the power ratio of two input polarization modes
and the phase difference $\varphi$. In Fig.$\,$\ref{fig:3}(b-c),
the transmission spectra of the H- and V-polarized modes are plotted
for different $\varphi$with $E_{v}=E_{H}=1$ and $\kappa_{H1}=\kappa_{V1}=1.5\kappa_{H2}=1.5\kappa_{V2}=1.5\kappa_{i}$.
When $\varphi=0$, the transmission spectra show coincident Lorentz
type and it is in critical coupling condition with $\kappa_{H,1}+\kappa_{V,1}=\kappa_{H,2}+\kappa_{V,2}+\kappa_{i}$.
There is a special situation that when $\varphi=\pi$, the transmission
of both the H- and V-polarized light is $1$. It seems that the MMR
is absent. The transmission spectra of the H- and V-polarized light
become asymmetrical Fano-lineshapes when $\varphi$ is neither $0$
nor $\pi$. As two polarization modes can convert to each other in
the MMR, which may induce the transmission to exceed 1. All of these
Fano-like lineshapes can be described by the equation of $T(\Delta)=\frac{(q\frac{\Gamma}{2}+\Delta)^{2}+b^{2}}{(\frac{\Gamma}{2})^{2}+\Delta^{2}}$
\cite{Gallinet2011An}. The asymmetry parameters $q$ and $b$ are
shape parameters describing the Fano-like interference. By fitting
the lineshapes with the equation, we can get the parameters $q$ of
each curve in Fig.$\,$\ref{fig:3}(b). Since changing the phase difference
$\varphi$ affects the transmission spectra, it can be used for sensing.
In addition, by solving the equations of $E_{out2}$ and $E_{out3}$,
we can get the power and the phase of two polarized light at the input
port, which means the device may also be used as a polarization analyzer. 

In fact, the MMR coupled with waveguides is a linear interference
structure, the independent H- and V-polarized light are two interference
channels. For a certain wavelength, when two polarization modes are
incident, rotating the MMR is equivalent to changing the relative
phase difference of two interference channels at input ports, which
will change the output power of two channels. However, when only one
polarization mode is input, changing the initial phase of the incident
light do not affect the time-averaged power of the output light.

For coupled MMRs, which can be treated as a more complex interference
structure, as shown in Fig.$\,$\ref{fig:3}(d). The radius of two
cavities are 100 $\mathrm{\mu m}$ and the conversion rate of the
PRs in MMRs is 1. The loss of the MMR1 is larger than that of the
MMR2. The transmission matrix method is used to analyze the influence
of different $\theta_{1}$ and $\theta_{2}$ on the transmission spectrum
with only H-polarized input. We found that the transmission spectra
for a specific polarization mode are the Lorentz lineshapes of MMR1
with a coupled-cavities-induced splitting as shown in Figs.$\,$\ref{fig:3}(e-f)
, which is similar to that in traditional coupled microrings. Due
to the coupling between two polarization modes in MMRs, whatever the
input light is, there are always two polarization modes in MMRs. According
to Fig.$\,$\ref{fig:3}(a-c), rotating MMR will change the power
coupling between MMRs, which means the intrinsic parameters of the
interference structure are changed, and thus the degree of the splitting
is affected. Meanwhile, rotating the MMR1 also equivalently changes
the initial phase of the input H-polarized light, but according to
the previous analyses, it will not affect the output power. So the
results of changing $\theta_{1}$ are similar to that of changing
$\theta_{2}$, as illustrated in Figs.$\,$\ref{fig:3}(e-f). Additionally,
such a coupled MMRs can be used for the polarization dependent sensing.
When there is an external perturbation to the MRR2, it will induce
round-trip phase shift to both polarization modes, i.e. $\phi_{h}$
and $\phi_{v}$. The summation of the two phases $\phi_{h}+\phi_{v}$
will induce a resonance frequency shift of MMR2, while the difference
of the two phases $\phi_{h}-\phi_{v}$ is equivalent to a geometric
rotation, i.e. the changing of $\theta_{2}$, thus the lineshape of
the spectrum will also be changed. 

In conclusion, a new type of microring resonator, which is analogous
to the M\"obius strip is proposed. The theoretical analyses reveal that
the \textquotedblleft M\"obius\textquotedblright{} microring resonator
holds many distinguishable properties, including the nontrivial mode
polarization properties and the fractional mode FSR compared to traditional
microring cavity of the same size. Besides, the spectral properties
of single MMR and coupled MMRs are studied. The spectra show Fano-like
features, which are related to the relative phase and the power ratio
of the input light with orthogonal polarizations. Such features might
be used in sensing and polarization analysis. Our work opens the door
towards the topological photonics by exploiting virtual dimension
of mode family and provides a novel way to engineering photonic structures
for functional devices and fundamental studies.
\begin{acknowledgments}
CLZ would like to thank X. Xiong, L. Jiang, H. X. Tang, X.-F. Ren
and Y. S. Zhao for discussions. This work was supported by the National
Key Research and Development Program of China (Grant No. 2016YFA0301300),
National Natural Science Foundation of China (Grants No. 61505195
and 11774110), and Anhui Initiative in Quantum Information Technologies
(AHY130000). The Open Project of State Key Laboratory of Advanced
Optical Communication Systems and Networks, Shanghai Jiao Tong University,
China (Grant No. 2018GZKF03002).
\end{acknowledgments}

\bibliographystyle{apsrev4-1}

\end{document}